\documentclass[prl,aps,twocolumn,floats,showpacs]{revtex4}
\usepackage{graphicx}
\usepackage{amsmath}

\newcommand{\be}{\begin{equation}}
\newcommand{\ee}{\end{equation}}
\newcommand{\bea}{\begin{eqnarray}}
\newcommand{\eea}{\end{eqnarray}}
\newcommand{\bse}{\begin{subequations}}
\newcommand{\ese}{\end{subequations}}

\newcommand{\lb}{\left[}
\newcommand{\rb}{\right]}

\newcommand{\lf}{\left\{}
\newcommand{\rf}{\right\}}

\renewcommand{\r}{\mathbf{r}}

\renewcommand{\H}{{\cal H}}

\newcommand{\E}{{\cal E}}

\renewcommand{\P}{{\cal P}}

\renewcommand{\S}{{\cal S}}
\newcommand{\tS}{\widetilde{\cal S}}

\newcommand{\tH}{\widetilde{\cal H}}
\newcommand{\tpsi}{\widetilde{\psi}}

\newcommand{\ts}[1]{\textstyle{#1}}

\newcommand{\half}{\textstyle{\frac12}}

\newcommand{\ket}[1]{\left.|  #1 \right>}

\begin{document}
\title{Transverse field effect in graphene ribbons}
\author{D. S. Novikov}
%% \email{dima@alum.mit.edu}
\affiliation{W. I. Fine Theoretical Physics Institute, 
University of Minnesota, Minneapolis, MN 55455}

\date{July 31, 2007}
% this is the resubmitted version: June 15, 2007

%\setlength\topmargin{0.0cm}

\begin{abstract}
It is shown that a graphene ribbon, a ballistic strip of carbon monolayer, 
may serve as a quantum wire whose electronic properties can be continuously
and reversibly controlled by an externally applied transverse voltage.
The electron bands of armchair-edge ribbons undergo dramatic transformations: 
The Fermi surface fractures, Fermi velocity and effective mass change sign,
and excitation gaps are reduced by the transverse field. 
These effects are manifest in the conductance 
plateaus, van Hove singularities, thermopower, and activated transport.
The control over one-dimensional bands
may help enhance effects of electron correlations, and be utilized
in device applications.
\end{abstract}


\maketitle
%%%%%%%%%%%%%%%%%%%%%%%%%%%%%%%%%%%%%%%%%%%%%%%%%%%%%%%%%%%%%%%%%%%%%%%%%%%%%

Building nanoscale systems with pre-determined properties 
has long been the focus of basic and applied research.
Progress in this field is tied to the recent advancements
%via the semiconductor 
in the synthesis of quantum nanowires \cite{lieber} 
and quantum dots \cite{bawendi} 
%with pre-determined properties 
via control of the growth process, 
as well as in the growth and selection of carbon nanotubes
\cite{Dresselhaus}. The characteristics of these devices, however,
are set by design and are typically difficult to modify. 
Ideally, one would like to be able to tune the system's properties 
reversibly after synthesis.
%It is desirable to be able to
%tune the system's properties reversibly after synthesis. 

In the present work we suggest that a graphene ribbon (GR),
a ballistic strip of recently discovered \cite{novoselov} carbon monolayer, 
may serve as a quantum wire whose electronic properties can be continuously
and reversibly controlled by the external transverse voltage.
The setup makes use of the massless relativistic 
electron dispersion in graphene \cite{novoselov,deheer,zhang}, with 
the valence and conduction bands touching at a conical Dirac point
\cite{wallace,Dresselhaus}.

Electron dispersion in GRs varies depending on their chirality, as the transverse 
confinement of Dirac fermions is sensitive to the boundary conditions
\cite{ribbons-dresselhaus,ribbons-fujita,aoki,ryu,ribbons-brey,ribbons-guinea,ribbons-ezawa,ribbons-scuseria,silvestrov-efetov}.
This has prompted proposals for GR applications 
as field-effect transistiors \cite{ribbons-appl}
and valley filters \cite{graphene-valley}.
Furthermore, GRs have been suggested as a host of interesting many-body phenomena,
including spin polarization on the edges 
\cite{ribbons-louie-nature,ribbons-louie-prl}, and as a basis for
building coupled electron spin qubits \cite{graphene-qubits}.
Recently spectral gaps in GRs have been measured  \cite{ribbons-exp},
scaling approximately inversely with the ribbon width.

The basic idea of the present proposal is that the properties 
at the Dirac point are fragile and can be affected by external fields \cite{lukose}.
Not suprisingly, the proposed strong electric field effect is similar
to that considered for carbon nanotubes \cite{nt-fet,ntanomaly,Rotkin}. 
Unfortunately, small radius $R\sim 1\,$nm of single-walled tubes 
requires very large transverse fields 
%$e\E R\sim \hbar v/R \sim 10\,$MV/cm 
$\E_{\rm [MV/cm]} \simeq 25/R_{\rm [nm]}^2$; 
this has so far hindered observation of band transformation.
%$\E \sim \hbar v/ e R^2 \sim 10\,$MV/cm. 
%This obstacle has so far hindered observation of these effects. 
Remarkably, with GRs, the ribbon width 
(that plays the role of the tube circumference) may vary in a broad range,
$L\sim 10-200\,$nm \cite{ribbons-exp}, and the effects 
of strong band transformation become realistic.

%%%%%%%%%%%%%%%%%%%%%%%%%%%%%%%%%%%%%%%%%%%%%%%%%%%%%%%%%%%%%%%%%%%%
\begin{figure}[b]
\includegraphics[width=3.5in]{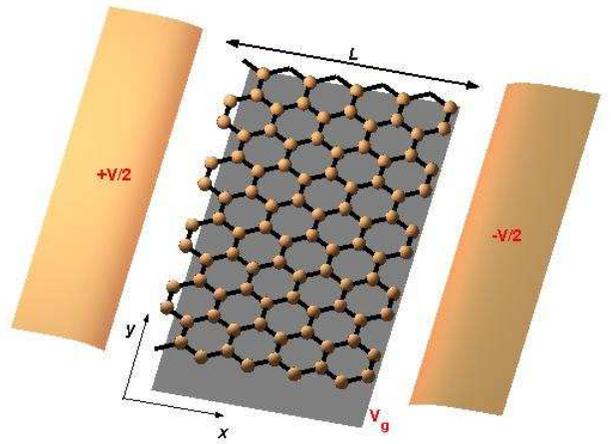}
\caption{(color online).
The setup. Left and right electrodes carry the voltage $\pm V/2$, 
producing the external field $\E_{\rm ext}$.
%$V \simeq \E_{\rm ext} L$, 
The carrier concentration is controlled by the back gate
voltage $V_g$.
  }
\label{fig:setup}
\end{figure}
%%%%%%%%%%%%%%%%%%%%%%%%%%%%%%%%%%%%%%%%%%%%%%%%%%%%%%%%%%%%%%%%%%%%    

%%%%%%%%%%%%%%%%%%%%%%%%%%%%%%%%%%%%%%%%%%%%%%%%%%%%%%%%%%%%%%%%%%%%
\begin{figure}[t]
\centerline{
\begin{minipage}[t]{3.5in}
\vspace{0pt}
\centering
\includegraphics[width=3.5in]{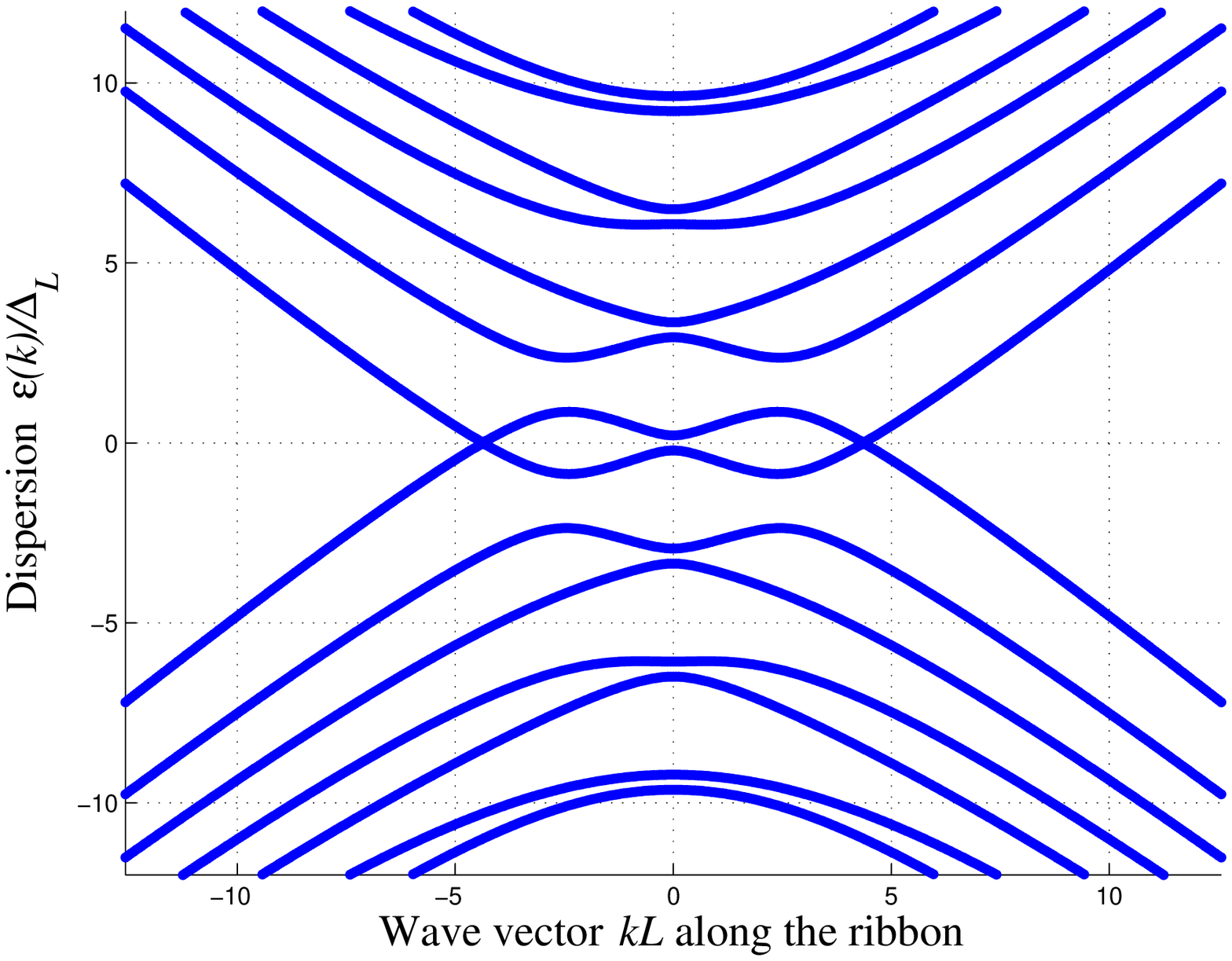}
\end{minipage}
\hspace{-1.4in}
\begin{minipage}[t]{1.3in}
\vspace{.035in}
\centering 
\includegraphics[width=1.3in]{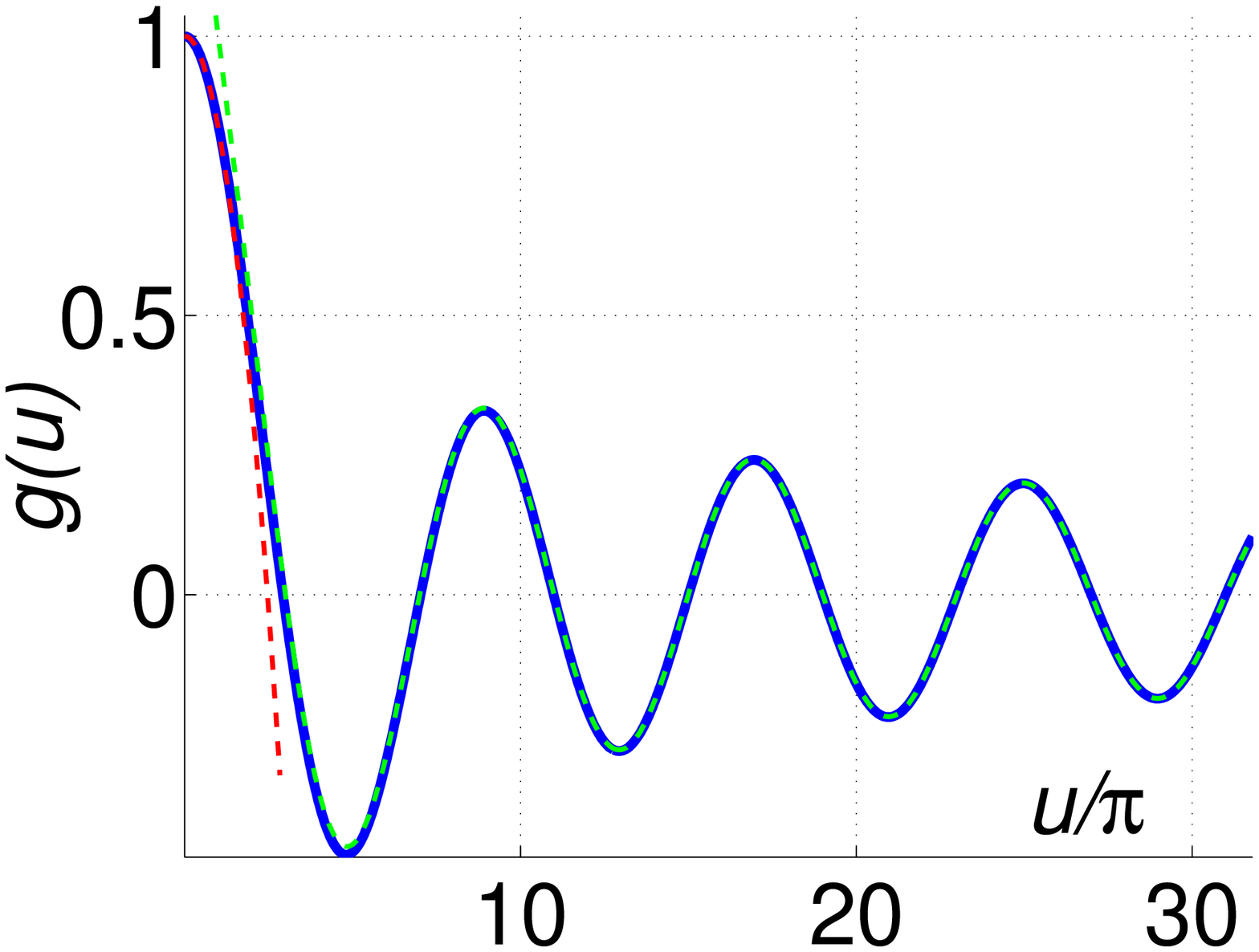}
\end{minipage}
}
\includegraphics[width=3.5in]{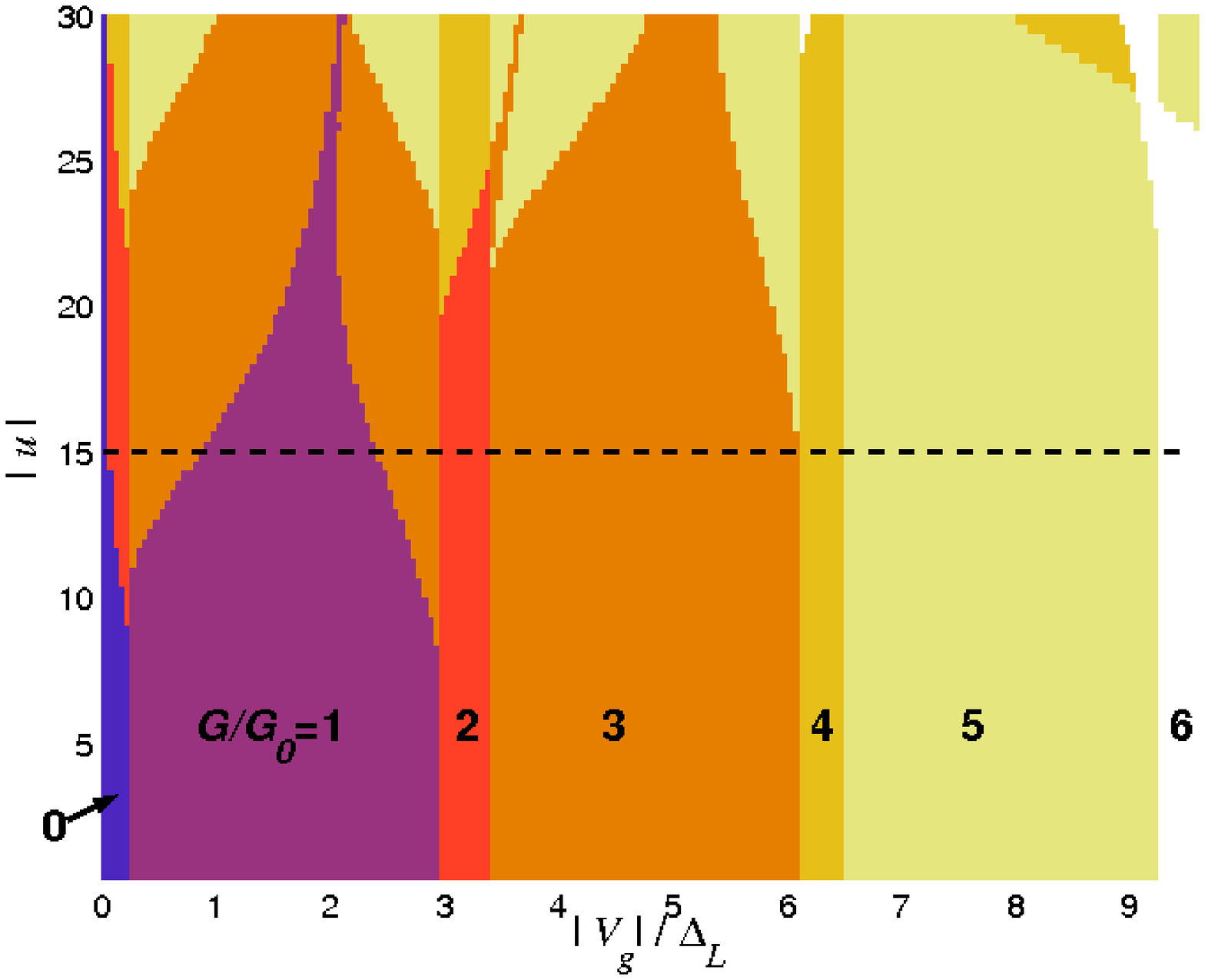}
\caption{(color online).
Transverse field effect in metallic GRs.
%$u=e{\cal E}L/\Delta_L \gg 1$. 
{\it Inset:} 
Velocity reversals in metallic GRs occur at the zeros 
of the function $g(u)$, Eq.~(\ref{g}); first reversal voltage $u_1\approx 9.2$.
Fine lines show the $u\ll 1$ and $u\gg 1$ asymptotic behavior of $g(u)$.
{\it Top:} The voltage $u=15$ above the reversal value $u_1$.
The Fermi surface acquires a pair of small pockets. 
Small gaps at $k=0$ are due to imperfect boundaries
\cite{ribbons-louie-prl}.
%{\it Inset:}
%Dipole moment $\P$ per one fermion species versus the field 
%$u=e{\cal E}R^2/\hbar v$, obtained as $\P=-dW/du$ with $W$ given by 
%(\ref{eq:Wintegral}). The weak field result obtained from 
%(\ref{Wlinearized}) is also shown.
%The cusp (marked) occurs at the field where
%velocity changes sign in metallic NT.
{\it Bottom:}
Landauer conductance $G$ (bold integers and colors)
in the units of $G_0=2e^2/h$, as the number of 
transverse modes at the Fermi energy $V_g$.
%controlled  by the gate voltage $V_g$.
Dashed cut corresponds to the top panel. 
%The field $u=27$ in-between the 2nd and 3rd velocity reversal points,
%with the Fermi surface breaking into two pairs of pockets. 
  }
\label{fig:disp-met}
\end{figure}
%%%%%%%%%%%%%%%%%%%%%%%%%%%%%%%%%%%%%%%%%%%%%%%%%%%%%%%%%%%%%%%%%%%%    
%%%%%%%%%%%%%%%%%%%%%%%%%%%%%%%%%%%%%%%%%%%%%%%%%%%%%%%%%%%%%%%%%%%%
\begin{figure}[t]
\centerline{
\begin{minipage}[t]{3.5in}
\vspace{0pt}
\centering
\includegraphics[width=3.5in]{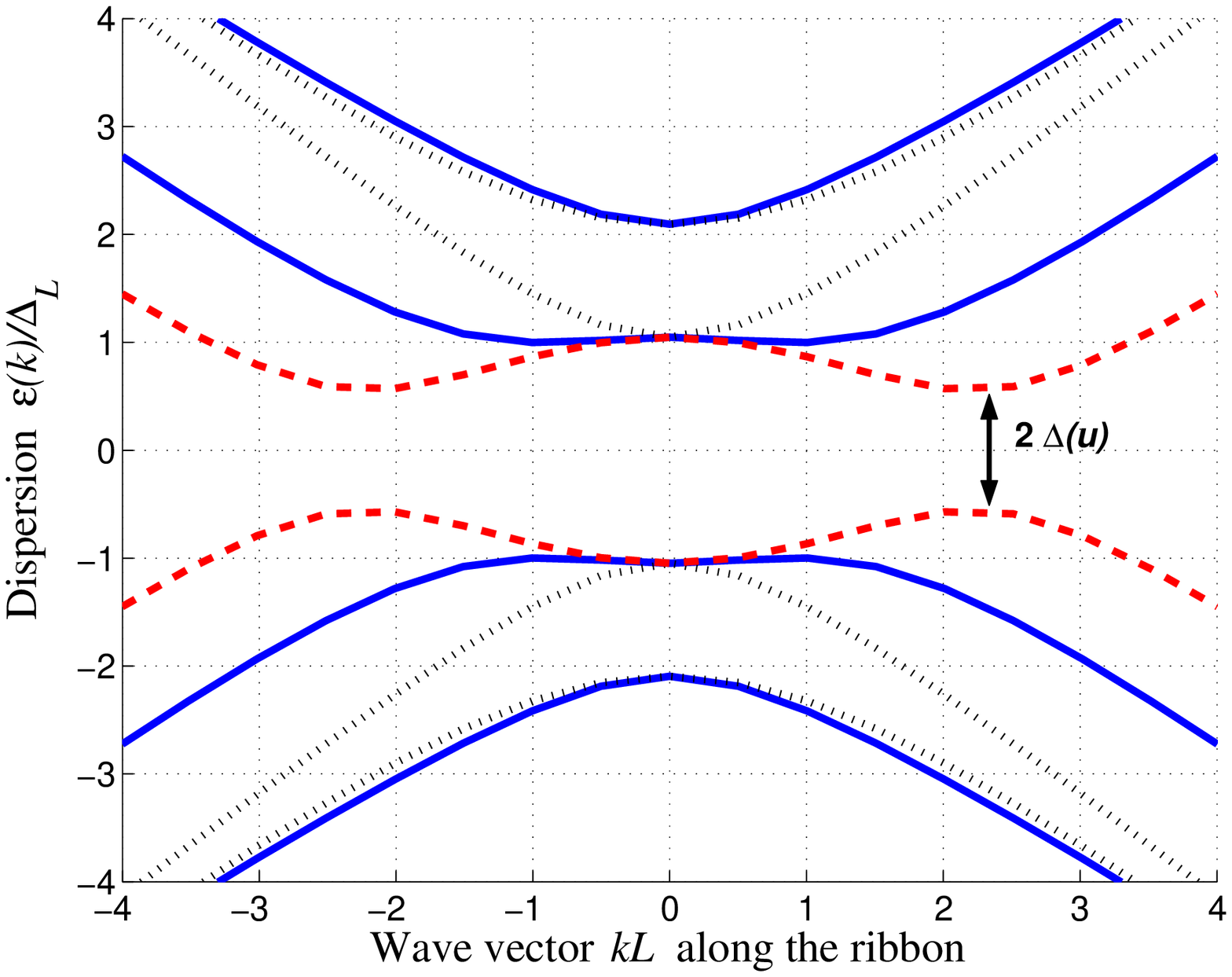} 
\end{minipage}
\hspace{-1.4in}
\begin{minipage}[t]{1.3in}
\vspace{.0035in}
\centering 
\includegraphics[width=1.3in]{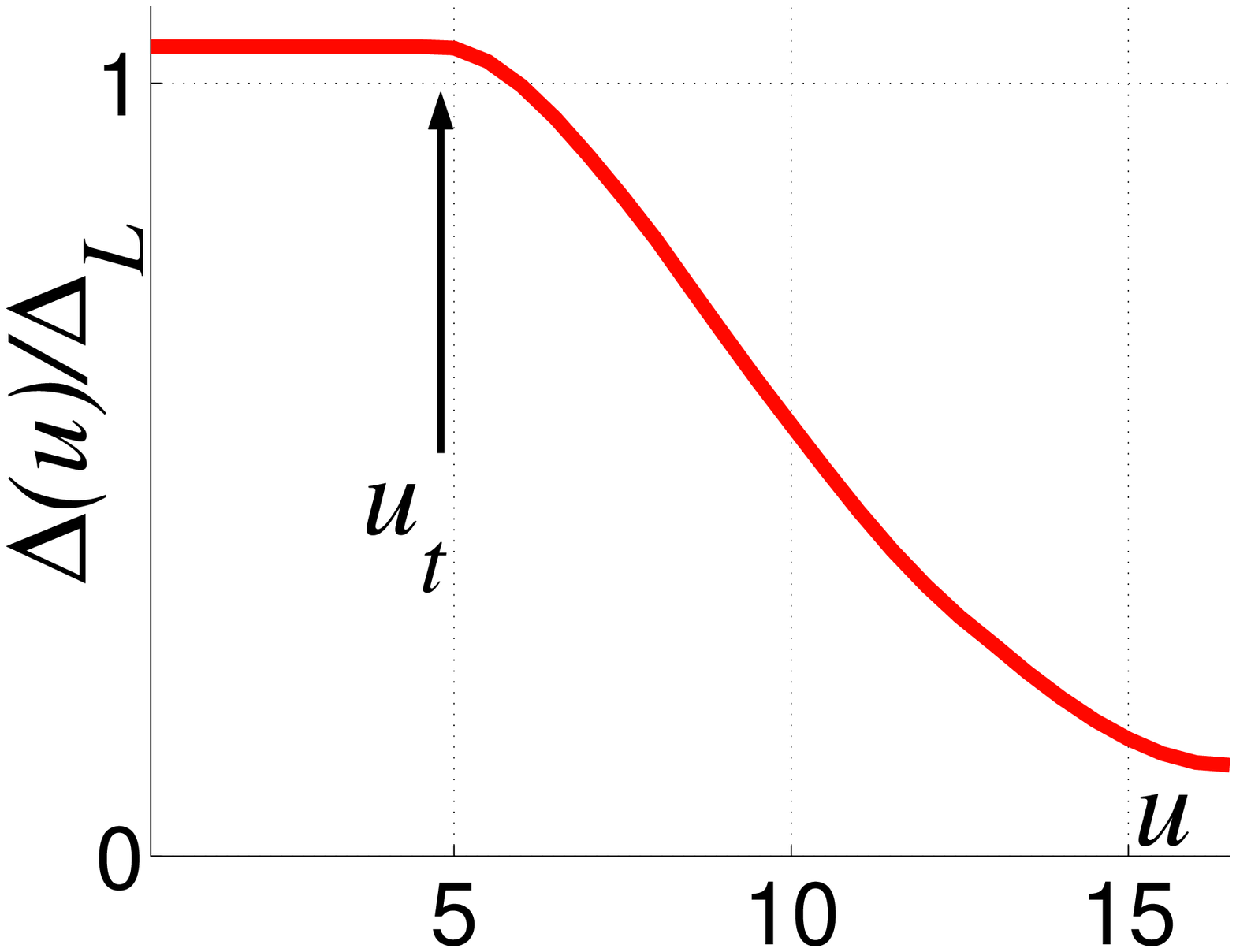}
\end{minipage}
}
\includegraphics[width=3.5in]{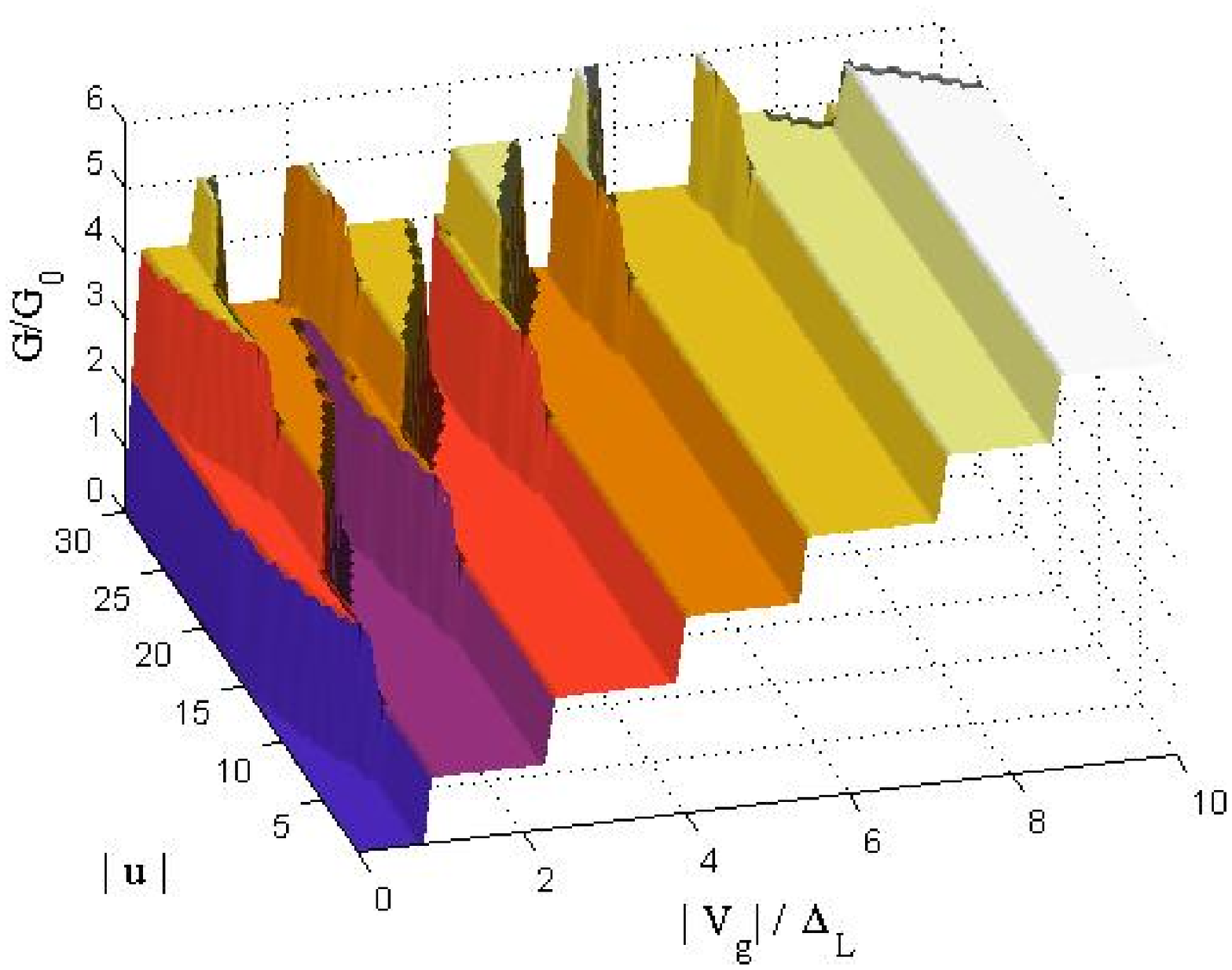}
%%%%%\vspace{-3mm}
\centerline{
\begin{minipage}[t]{3.3in}
\vspace{0pt}
\centering
\end{minipage}
\hspace{-1.5in}
\begin{minipage}[t]{1.4in}
\vspace{0pt}
\centering 
\end{minipage}
}
\caption{(color online).
Transverse field effect in semiconducting GRs. 
%GRs in the strong field-effect regime,
%$u=e{\cal E}L/\Delta_L \gg 1$. 
For voltages $u>u_t\approx 4.5$ above threshold
the effective mass at $k=0$ is negative.
{\it Top:} 
Bands transformation, starting from $u=0$ (dotted), to just above  
threshold $u=6$ (solid), to above threshold, $u=10$ 
(dashed, only lowest subbands shown), where the 
gap is minimal at $k\neq 0$.
{\it Inset:} 
Gap suppression occurs above the threshold $u_t$.
{\it Bottom:}
Landauer conductance plateaus in the units of $G_0=2e^2/h$.
%as a function of the transverse voltage $u$
%and the gate voltage $V_g$.
%For very large fields, the band transformation away from $k=0$
%remind those in metallic GRs. 
%Here the field $u=27$, the features at $kL\sim 10$ are similar to the 
%bottom part of Fig.~\ref{fig:disp-met}.
  }
\label{fig:disp-sc}
\end{figure}
%%%%%%%%%%%%%%%%%%%%%%%%%%%%%%%%%%%%%%%%%%%%%%%%%%%%%%%%%%%%%%%%%%%%    

In the setup shown in Fig.~\ref{fig:setup}, electrons in a GR 
are confined along the $x$ axis, 
while the longitudinal momentum $k_y\equiv k$ is conserved.
The effect of the applied transverse voltage $V$ is to induce the potential
%results in the external potential that 
\be \label{U}
U(x) = - e\E (x-L/2) \,.
\ee
%For the Fermi energy close to the band center, graphene is a semimetal,
The acting field $\E\propto \E_{\rm ext}\propto V$ 
can be assumed uniform and proportional to the external field $\E_{\rm ext}$
as long as the bands are not strongly mixed 
(as described below); $e$ is the unit charge.
We subtracted the average, setting $\int_0^L \! dx\, U(x) = 0$
(the subtracted constant adds to the chemical potential controlled 
by the gate voltage $V_g$).    
The natural units for 
%momentum, $\bar k = kL$, 
the transverse voltage %$\Delta_L$
%$\bar \epsilon = \epsilon/\Delta_L$, 
and energy 
\be \label{u}
u={e\E L/\Delta_L}\,, \quad 
\Delta_L = {\hbar v/L} \simeq {0.7\,}{\rm eV}/L_{\rm [nm]} \,,
\ee
where $v\simeq 10^6\,$m/s is graphene's Fermi velocity.
The ballistic limit of transport is implied.

We now give an overview of the results. 
The electron band transformation %by means of the external voltage 
is shown in Figs.~\ref{fig:disp-met} and \ref{fig:disp-sc}. 
The longitudinal electron bands change qualitatively when 
the dimensionless transverse voltage approaches $u\sim 10$.
A number of effects follow:

(i) The Landauer conductance \cite{Landauer}
is quantized in the units of $G_0=2e^2/h$,
similar to that in point contacts in GaAs \cite{steps}.
%(factor of two is due to electron spin).
The crucial difference is that now the positions and widths of the plateaus
can be controlled by the transverse voltage.
The sharp steps in Figs.~\ref{fig:disp-met} and \ref{fig:disp-sc}
in real systems will be smoothened out by finite temperature or weak disorder, while 
the conductance values on the plateaus will remain equal to the quantized values.
%(provided the disorder is weak).

(ii) The thermopower $S\propto -\partial \ln G/\partial V_g$, 
being proportional to the conductance derivative \cite{thermopower}, peaks at the 
borders between the 
domains in Figs.~\ref{fig:disp-met} and \ref{fig:disp-sc} (bottom).

(iii) The Fermi velocity in metallic GRs is reduced by the field,
$v\to vg(u)$ [Fig.~\ref{fig:disp-met} inset and Eq.~(\ref{g}) below]. 
%Fig.~\ref{fig:disp-met} demonstrates that the effect of the strong field 
%may be quite dramatic. 
As a consequence, the one-dimensional density of states at the band center
$\nu(0) =2/\{\pi \hbar v |g(u)|\}$ increases
[factor 2 is due to spin degeneracy]. 
This increase magnifies the effects of electron interactions. 
The latter may manifest themselves via the increase of the Luttinger 
liquid exponent in a sufficiently long ribbon, and through excitonic instabilities 
(resulting in interaction-induced gaps).

(iv) The Fermi velocity changes sign for the field 
values corresponding to zeroes of $g(u)$, 
causing strong van Hove singularities in metallic GRs. 
The Fermi surface fractures, 
with each sign change adding a pair of small pockets to the Fermi surface
(Fig.~\ref{fig:disp-met}, top). This effect produces extra conductance plateaus 
(Fig.~\ref{fig:disp-met}, bottom).

(v) There is a threshold voltage $u_t\simeq 4.5$ above which
the effective mass of the lowest energy subband in semiconducting GRs changes sign,
so that the longitudinal electron dispersion acquires 
symmetric minima at small but nonzero $k$ (Fig.~\ref{fig:disp-sc}).
%At $u\sim 10$ the gap is closed noticeably. 
The excitation gap is then reduced by the field (Fig.~\ref{fig:disp-sc} inset).
This effect can be detected in the shift of 
the conductance plateaus (Fig.~\ref{fig:disp-sc}, bottom), and in the activated 
transport measurements.

(vi) The band structure remains electron-hole symmetric at any field 
for both metallic and semiconducting GRs due to the Dirac symmetry of the problem.
Thus the conductance plots of Figs.~\ref{fig:disp-met} and \ref{fig:disp-sc}
are independent of the polarity of the gate and transverse voltages.

Turning to possible applications,
the setup may serve as a field-effect transistor with a tunable working point,
in which the ``transverse'', $V$, and the ``normal'', $V_g$, 
field effects can be utilized separately.
Furthermore, one may selectively 
amplify combinations $\alpha (V-V^0) + \beta (V_g-V_g^0)$, 
$\alpha = \partial G/\partial V|_{\cal V}$ 
and $\beta=\partial G/\partial V_g|_{\cal V}$
by choosing an appropriate working point ${\cal V}=(V^0, V_g^0)$
%or $\delta V-\delta V_g$ depending on the working point.
on the edge of the conductance plateau.
Tuning the parameters to achieve a large gain for 
say, $V-V_g$, combined with the device's large input and low output impedance,
is reminiscent of an operational amplifier. 
By the same token, strong conductance 
nonlinarity in both inputs $V$ and $V_g$ may 
render this setup into a few-nm size signal multiplier, or even into a logic gate.

We now outline the details of the calculation.
At the $\pi$-band center ($\epsilon=0$), 
the electron dispersion is determined by the 
two inequivalent Dirac points in the Brillouin zone. 
The low-energy states
$\Psi(\r) = e^{iKx} \psi_+(\r) + e^{-iKx}\psi_-(\r)$
are represented \cite{Dresselhaus} 
in terms of the smoothly varying envelope $\psi = \{\psi_+, \psi_-\}$
that consists of the pair of the two-component spinors $\psi_+$ and $\psi_-$ 
with values on the two sublattices of the honeycomb lattice
[here $K=-4\pi/3a_0$, where $a_0=\sqrt{3}a_{cc}$ is the graphene lattice constant, 
and $a_{cc}=0.144\,$nm is the Carbon bond length].
The dynamics of the envelope is governed by the Dirac equation 
$\H\psi=\epsilon\psi$, with the effective Hamiltonian 
\be \label{H}
\H = \begin{pmatrix} \H_+ & 0 \\ 0 & \H_- \end{pmatrix},
\quad \H_\pm = \pm i\hbar v\sigma_1 \partial_x - \hbar vk \sigma_2 + U(x) \,,
\ee
where $\sigma_{1,2}$ are the Pauli matrices.
The boundary conditions $\Psi(\r)|_{x=0,L}=0$ at the armchair edges
dictate \cite{ribbons-brey} 
\be \label{BC}
\psi_+(0)+\psi_-(0)=0\,, \quad \psi_+(L)+ e^{i\phi_n}\psi_-(L)=0\,,
\ee
where the phase
%\be \label{phi}
$\phi_n = KL = -\ts{2\pi\over3}(n+1-\delta)$,
%\quad \mbox{and} \quad 
and $L=\half(n+1-\delta)a_0$  
%\ee
is the effective ribbon width (the distance between the sites on which 
$\Psi$ vanishes). The phase $\phi_n$ may incorporate corrections
coming from imperfect edges, 
similar to the curvature-induced corrections in nanotubes \cite{curvature}.
(For example, the $\delta t/t \approx 0.12$ 
change in the hopping amplitude at the edges 
due to the passivated bonds \cite{ribbons-louie-prl}
reduces the effective width $L$ and the boundary phase $\phi_n$
by the amount $\propto\delta = {3\sqrt{3}\over \pi} {\delta t\over t} \approx 0.20$.)

The system (\ref{H}) and (\ref{BC}) is solved numerically 
(Figs.~\ref{fig:disp-met} and \ref{fig:disp-sc})
via the transfer matrix approach
similar to that of Refs.~\cite{nt-fet,ntanomaly,Lee-Novikov}.
Eq.~(\ref{H}) is equivalent to 
%$\partial_x \psi_{\pm} = \pm \P\psi$, 
%and $\partial_x \psi' = -\P \psi'$, 
\be \label{P}
\partial_x \psi_{\pm} = \pm \P\psi_\pm \,, 
\quad \P(x) = k\sigma_3 + i\sigma_1 (U-\epsilon)/\hbar v \,.
\ee
The armchair boundary conditions (\ref{BC}) require
%\be \label{trSS}
$\mbox{tr\,} (\S \tS) = 2 \cos \phi_n $ 
%\ee
for the product of the transfer matrices
\be
\S = {\cal T}_x e^{\int_0^L \P(x)dx}\,, \quad
\tS = \widetilde{\cal T}_x e^{\int_0^L \P(x)dx} \,,
\ee
where ${\cal T}_x$ and $\widetilde{\cal T}_x$ symbolize the ``chronological''
and ``anti-chronological'' orderings of the operators $\P(x)$ 
that do not commute for different $x$.

In the absence of the field, the GR spectrum consists 
of one-dimensional Dirac bands
with $|\epsilon_{k=0}| = \Delta_L \times \frac\pi3 |n+1-3p - \delta|$, 
$p = 0, \pm 1, \pm 2, ...$ . Thus GRs with $n=3p-1$ are metallic
(with small gap $\propto \delta$ originating from imperfect boundaries 
\cite{ribbons-louie-prl}), in which case the lowest energy mode is non-degenerate,
and the rest are doubly-degenerate (the latter degeneracy is lifted 
by the finite $\delta$).
The ribbons with $n=3p$ and $n=3p-2$ are semiconducting,
with non-degenerate bands, and excitation gaps 
$|\epsilon_{k=0}| =\Delta_L\times \frac\pi3 (1\mp\delta)$ correspondingly.
%In what follows we consider perfect armchair edges 
%(i.e. set $\delta=0$) for simplicity  
%as it does not qualitatively affect our findings.

%{\it Metallic GRs.---}
%The main effect of the field in metallic ribbons 
%is to fracture the Fermi surface by changing the 
%Fermi velocity sign at $k=0$, as shown in Fig.~\ref{fig:disp-met}. 
%This can be understood by the means of the chiral gauge transformation
To study the transverse field effect it is convenient to employ 
the chiral gauge transformation \cite{nt-fet,ntanomaly}
\be \label{gauge}
\psi_\pm = e^{\pm i\sigma_1 \varphi(x)} \tpsi_\pm \,, 
\quad \varphi = \int_0^x \! U(x')dx'/\hbar v \,,
%\quad 
%\psi' = e^{-i\sigma_1 \int^x \! Udx/\hbar v} \tpsi 
\ee
that preserves the boundary conditions (\ref{BC}) and transforms 
the system (\ref{H}), $\H_\pm \to \tH_\pm$,
\be \label{H-gauged}
\tH_\pm = 
\hbar v\lb -k \sigma_2 e^{\pm 2i\sigma_1 \varphi(x)} \pm i\sigma_1 \partial_x \rb.
%\quad
%\tH' = \hbar v\lb -k \sigma_2 e^{-2i\sigma_1 \phi(x)} - i\sigma_1 \partial_x \rb .
\ee
The transformation (\ref{gauge}) shows that the spectrum at $k=0$  
is unaffected by the field. 
For the metallic GRs with ideal edges 
%corresponding to the boundary condition phase 
($e^{i\phi_n}\equiv 1$), 
the two degenerate $k=0$, $\epsilon=0$ eigenstates, each consisting 
of a pair $\{\tpsi_+, \tpsi_-\}$, are  
\[ 
\ket{1}=\frac12 \lf  
\begin{pmatrix} 1 \\ 1\end{pmatrix}, 
\begin{pmatrix} -1 \\ -1\end{pmatrix} \rf 
\ \mbox{and} \ \
\ket{2}=\frac12 \lf  
\begin{pmatrix} 1 \\ -1\end{pmatrix}, 
\begin{pmatrix} -1 \\ 1\end{pmatrix} \rf. 
\]
Projecting the Hamiltonian (\ref{H-gauged}) onto these states,
$\tH \to \hbar v k g(u)\sigma_2$, we find the spectrum around $k=0$ 
\be \label{g}
\epsilon=\pm \hbar v |kg(u)|\,, \quad
g = \int_0^1\! d\xi \, \cos \lb u\xi(1-\xi)\rb  \,.
\ee
The function $g(u)$ is plotted in Fig.~\ref{fig:disp-met} inset.
For $|u|\ll 1$, $g\simeq 1-u^2/60$.
Its $|u|\gg1$ form $g \simeq \sqrt{\pi/|u|} \cos[(|u|-\pi)/4]$
determines the successive voltages $u_n \approx \pm (3+4n)\pi $, $n=0,1,2,...$,
where the $k=0$ velocity changes sign. 
At those voltages the dispersion $\epsilon\sim k^3$ at $k=0$, causing
the van Hove singularity $\nu(\epsilon)\sim |\epsilon|^{-2/3}$ 
in the density of states at $\epsilon=0$, 
and an additional  pair of pockets of Fermi surface emerges.
In the $|u|\gg 1$ limit, such pockets appear at the zeroes of $g$
%for {\it any} boundary phase $\phi_n$ (i.e. 
for both metallic and semiconducting GRs.

Electron interactions in graphene result in the RPA screening 
of the external field $\E_{\rm ext}$. The screening is scale-invariant,
$\E = \E_{\rm ext}/\kappa$, for an infinite sheet
\cite{graphene-pol}, $\kappa = 1+2\pi e^2 /4\hbar v\simeq 5$. 
The depolarization problem in nanotubes 
\cite{Benedict95,nt-fet,ntanomaly,Rotkin,Krcmar,Brothers05,Kozinsky}
also yields $\kappa\simeq 5$ practically independent of the tube radius
and chirality. This linear-screening estimate will remain valid in GRs 
as long as the subbands are not strongly mixed. 
In the opposite case the field on the ribbon edges, estimated in the Thomas-Fermi
fashion, develops an algebraic singularity \cite{chklovskii},
which corresponds to filling the Fermi-surface pockets. 
%this field nonuniformity will affect the number of mixed subbands.
%does not qualitatively affect our findings.
%(otherwise, the field at the edges will develop a singularity \cite{chklovskii}). 
As a result, the uniform field
model (\ref{U}) is justified for weak to moderate fields.
For the acting field $u=10$, the required external field 
$u_{\rm ext}\simeq 50$ 
is achieved at $\E_{\rm ext} L \simeq 1$\,V across the ribbon width $L=30\,$nm.

To conclude, the transverse voltage applied across a graphene 
nanoribbon dramatically affects its longitudinal electronic dispersion.
The Fermi surface breaks up into pockets for the metallic ribbons,
and the excitation gap closes for the semiconducting ones.
The strong field effect can lead to interesting physical phenomena 
as well as be utilized in carbon-based electronic devices.

This work has benefited from illuminating discussions with M. Fogler, 
L. Glazman and L. Levitov. 
The research was sponsored by NSF grants DMR 02-37296 and DMR 04-39026.

%%%%%%%%%%%%%%%%%%%%%%%%%%%%%%%%%%%%%%%%%%%%%%%%%%%%%%%%%%%%%%%%%%%%

\end{document}